\documentclass{article}
\usepackage[T1]{fontenc} 
\usepackage[utf8]{inputenc} 
\usepackage{ismir,amsmath,cite,url}
\usepackage{graphicx}
\usepackage{color}

\usepackage{lineno}
\usepackage[symbol]{footmisc}

\title{Improving Polyphonic Music Models with Feature-Rich Encoding}

\oneauthor
{Omar Peracha} {Humtap (research conducted independently) \\ {\tt omar.peracha@gmail.com}}

\def\authorname{Omar Peracha}

\begin{document}

\maketitle
\begin{abstract}
This paper explores sequential modelling of polyphonic music with deep neural networks. While recent breakthroughs have focussed on network architecture, we demonstrate that the representation of the sequence can make an equally significant contribution to the performance of the model as measured by validation set loss. By extracting salient features inherent to the training dataset, the model can either be conditioned on these features or trained to predict said features as extra components of the sequences being modelled. We show that training a neural network to predict a seemingly more complex sequence, with extra features included in the series being modelled, can improve overall model performance significantly. We first introduce TonicNet, a GRU-based model trained to initially predict the chord at a given time-step before then predicting the notes of each voice at that time-step, in contrast with the typical approach of predicting only the notes. We then evaluate TonicNet on the canonical JSB Chorales dataset and obtain state-of-the-art results.
\end{abstract}
\section{Introduction}\label{sec:introduction}

Computational modelling of polyphonic music is now a decades-old practice, with documented attempts dating back over sixty years [1]. Recent years have seen great progress in these models' ability to capture the semantics of a musical corpus, with much of this progress due to advances in artificial neural network algorithms and their application to the music domain.

Some of the most significant breakthroughs of late have experimented with applying newly-developed architectures to the problem of modelling symbolic music [2-3], training or pre-training on a large cross-domain corpus [2][4], or introducing Gibbs-like sampling methods to orderless models [5-6]. We instead focus on the sequence being modelled itself, and provide observations and enhancements that might improve results across a wide range of approaches. 

We conduct experiments with two architectures: a multi-layer Transformer encoder [7] with input masking and a model based on the Gated Recurrent Unit [8] to which we give the nickname TonicNet. Using the JSB chorales dataset, split into training, validation and test sets as per [9], the models are trained to predict each token of the samples one-by-one in a sequential manner. 

We observe that increasing the amount of musical information these models are trained to predict tends to improve overall performance of both models as measured by validation set loss. In particular, by training to first predict the chord at a given time step before then predicting the notes of each voice at that time step, both models show improvements in validation loss, despite the modelled sequences being longer and containing a larger possible output space than would be the case if predicting only the notes. We also train TonicNet on smaller subsets of the samples by restricting the number of voices being modelled, and again observe that results improve when more musical information is contained in the sequence being predicted.

These observations allow better results to be achieved without the need to use models with an ever-larger number of parameters. Concretely, TonicNet is approximately an order of magnitude smaller than Music Transformer, yet obtains a lower validation set loss on the JSB chorales dataset than that reported by [3] as a result of being trained to predict both chords and notes. It also generates samples much faster than reported in [5] by avoiding Gibbs-like sampling, and achieves state-of-the-art validation loss without requiring pre-training on a larger cross-domain corpus, saving a significant amount of training time.

All code for this paper is made publicly available, including the ability to load and sample from the pre-trained TonicNet model.\footnote{ {\tt https://github.com/omarperacha/TonicNet} }
Samples generated by TonicNet are also made available in both MIDI and audio form.

\section{Related Work}
Recurrent Neural Networks have been noted for their ability to model sequential data, with the LSTM [10] in particular being a favoured approach for a number of applications. RNNs have been widely-used in recent attempts to model musical data, which lends itself well to a sequence-based representation. The earliest such examples used RNNs to model monophonic music [11-12].

In [9] RNNs were combined with Restricted Boltzmann Machines to model polyphonic music, while RNNs and LSTMs were combined with Deep Belief Networks for the same task by [13] and [14] respectively. BachBot [15] also uses an LSTM-based neural network, and both the architecture and hyperparameters described in their paper are influential in this work. All of these polyphonic models have been evaluated on the Bach chorales, although not all with the same dataset split. BachBot is trained to predict not only the notes, but also whether a fermata, the symbol used by Bach to mark phrase endings, coincides with each given note. This is deemed to improve sample quality, however the version of the dataset used in this work does not contain information regarding presence of fermatas.

The Transformer and its derived architectures have been popular choices for more recent polyphonic music models, and are emerging as strong alternatives to RNNs more broadly due to the self-attention mechanism demonstrating great effectiveness at capturing long-term dependencies in training data [7]. Music Transformer adds a novel relative attention mechanism to the vanilla Transformer and is evaluated on the JSB chorales dataset. LakhNES [4] uses the Transformer-XL architecture [16] and is pre-trained on a large corpus of four-part music before being fine-tuned on the NES Music Database [17]. An event-based encoding is preferred which allows for more precise rhythm than the most commonly seen approach of slicing the input sample along a 16th-note (or some other suitable, dataset-dependent value) grid. MuseNet [2] is based on the GPT-2 Transformer model [18] and is trained on a massive corpus of polyphonic music. The data representation used has similar qualities to the event-based representation described by [4], and allows the trained model to sample while taking into account specific instrumentation and musical style. The encoding also includes information relating to note loudness.

Chords have been used in some previous approaches to improve the quality of monophonic music models. Chords are used as extra inputs to aid the generation of melodies by [19], who use a dual LSTM network Product of Experts system [20], and by [21], who propose training a convolutional generative adversarial network to generate melodies one bar at a time. However, neither of methods learn to predict the chords being used as input for each proceeding step, but rather provide them as fixed inputs.

HARMONET [22], comprises an ensemble of multiple neural networks trained with the ultimate goal of harmonising a given melody in the style of Bach. The first step is to derive the chords at each quarter-note step from the melody, relative to the key, including the inversion (and therefore the bass part). The inner parts are then predicted in a second step. 

In the method proposed in this paper, chords are in fact included in the sequence as a de facto extra voice which must be predicted along with the other four voices, rather than being used as a secondary conditioning input. Inversions are not encoded in the chord representation, as the bass part is included later in the sequence.

DeepBach [5] is also trained on a representation including extra musical features as added voices. In this case, it is the presence of a fermata at each time step. The model is not trained to predict this, however, but fermata information is instead provided as a fixed input which helps guide the musical structure of the generated samples. Chord tokens are not used by the DeepBach model.

Both DeepBach and COCONET [6] are trained with the primary goal of completing partially-filled musical scores, for example harmonising a given melody, though both are able to generate entire four-voice samples from an empty or randomly-initialised score. They do both require the length of the sample in time-steps to be preset in order to facilitate the orderless Gibbs-like sampling methods used, and DeepBach further requires fermata information to be provided. TonicNet instead uses ancestral sampling to generate scores, and is trained to predict successive tokens in a purely autoregressive fashion, requiring absolutely no preset information relating to length or phrasing. This ultimately inhibits sample quality as the phrase lengths may not display the consistent symmetry over time observed in the training corpus, however the model still obtains a lower validation set loss on the JSB chorales dataset than the upper bound reported by [3] for orderless evaluation of COCONET (i.e. evaluating COCONET's ability to correctly fill in the missing notes in partially-completed scores), when averaging purely over the note predictions and ignoring the chords which do not originally appear in the benchmark dataset.

A powerful advantage of DeepBach and COCONET is that they lend themselves far more naturally to interactive applications. Theoretically one could fix the chords or the notes of a given voice when sampling from TonicNet by ignoring predicted output for the relevant part, instead using the token from the corresponding time-step of the fixed sequence as input to the model at the next time-step. However, this kind of forced sampling has not been tested empirically and so sample quality under these conditions cannot be attested.

\section{Dataset}\label{sec:dataset}

\subsection{JSB Chorales}\label{subsec:jsb}

The JSB chorales are the most commonly-used benchmark for measuring the performance of polyphonic music models to date. They are a set of short, four-voice pieces well-noted for their stylistic homogeneity. The chorales were originally composed by Johann Sebastian Bach in the 18th century. He wrote them by first taking pre-existing melodies from contemporary Lutheran hymns and then harmonising them to create the parts for the remaining three voices. The version of the dataset used here consists of 382 such chorales, with a train/validation/test split of 229, 76 and 77 samples respectively.\footnote{ The version of the dataset used can be accessed at {\tt https://github.com/czhuang/JSB-Chorales-dataset} }

\subsection{Representation}\label{subsec:representation}

\subsubsection{Serialisation}

The dataset is pre-serialised onto a 16th-note grid, which captures full resolution of the original chorales. Only the pitch information of the four voices at each time-step is encoded in the canonical dataset; other symbolic data that may appear in a musical score, such as loudness or fermata, are absent. Furthermore, information regarding note boundaries in the case of repeated pitches is not present. The consequence of this is that it is not possible to truly accurately model Bach's original rhythms, unlike other approaches, e.g. [5], where the version of the dataset used allows note boundary information to be preserved using a special 'hold' token, despite 16th-note serialisation. A partial workaround to lack of repeated note boundaries during sampling is to simply tie together consecutive occurrences of the same pitch in a voice, which we refer to as rhythmic 'smoothing', however this inevitably sacrifices some rhythmic integrity of the original chorales. While it may seem unideal to serialise music onto a fine-resolution rhythmic grid, as this has the consequence of vastly extending the length of the sequences being modelled, it in fact acts as a highly effective form of data augmentation; brief experiments which encoded the true duration values of the notes, using the Bach chorales as made available by the music21 toolkit [23], ultimately performed significantly worse than the representation presented in this work, both in terms of validation loss and sample quality.

We include chords in our encoding. We first derive the chords for each 16th-note time-step by analysing the pitches of the four voices at said time-steps, using the music21 chord module. We then create a single ordered sequence for each sample in the form C\textsubscript{0}, S\textsubscript{0}, B\textsubscript{0}, A\textsubscript{0}, T\textsubscript{0}, C\textsubscript{1}, S\textsubscript{1}, B\textsubscript{1}, A\textsubscript{1}, T\textsubscript{1}... C\textsubscript{N-1}, S\textsubscript{N-1}, B\textsubscript{N-1}, A\textsubscript{N-1}, T\textsubscript{N-1}, <END>, where C, S, A, T and B represent the Chord, Soprano, Alto, Tenor and Bass inputs at each respective step, and N is the total number of 16th-note time-steps in the given sample. The model is fed the elements of the sequence one-by-one and tasked with predicting the next element in each case, thus it must predict the chord governing each 16th-note time-step before then predicting the actual pitch values observed in the dataset at that time-step. The effect of predicting the Bass note before the Alto and Tenor notes seems to be negligible, but has not been thoroughly tested and so may be considered arbitrary. The longest sequence observed in the dataset given the described representation, is 2,881 tokens in length, taking into account the appended <END> token, but the lengths vary quite considerably across the dataset.

\subsubsection{Symbolic Encoding}

Pitches and chords are all represented by distinct integer values. We restrict pitches purely to those observed in the dataset (MIDI values 36-81 inclusive assuming a value of 60 to be Middle C). The MIDI pitch value 37, despite not in fact appearing in the dataset, is included so as to better facilitate data augmentation by transposition, described in the next section. Chords are represented as belonging to one of 50 classes, comprised by 12 major chords (one chord per pitch class in the western chromatic scale), 12 minor chords, 12 diminished chords, 12 augmented chords and a special <OTHER> token which accounts for any chord which is not interpreted as fitting into the previous 48 classes. A <CHORD REST> token completes the set used to represent chord classes, and denotes instances when all four voices have rests at that time-step. Any voice-wise occurrence of a rest in the dataset is itself represented by a distinct <REST> token, and finally the <END> token completes the set of possible model input/output classes, taking the total to 98.

\subsection{Conditioning}\label{subsec:conditioning}

In addition to chord/pitch inputs, which we refer to as X-input, we condition the model a second input that relays information about note repetition. Concretely, alongside each X-input value X\textsubscript{\textit{n,i}}, where \textit{i} is used to index the chords and voices, we also input an integer, Z\textsubscript{\textit{n,i}}, corresponding to the number of consecutive times the value represented by X\textsubscript{\textit{n,i}} has so far appeared in voice \textit{i}, resetting to 0 each time a new value is observed in that voice or if Z\textsubscript{\textit{n,i}} exceeds 79 (equating to 5 bars in 4/4 timing). This value was chosen because only one sample features Z-input values greater than 79, with this sample's maximum Z-input value of 143 being a clear outlier.

The motivation for this is that we might more explicitly capture some of the inter-voice rhythmic relationships that exist in the music, and indeed we observe that it improves model performance (Table 1). We also experimented with instrument labelling, as in [3], and found that while it somewhat improved model performance, repetition encoding had a more significant impact and combining both repetition encoding and instrument labels did not perform better than repetition encoding alone. From this we can infer that repetition encoding, as well as helping the model to better learn timing information relating to note and chord changes, also fulfils a role similar to instrument labelling. We refer to repetition encoding inputs as Z-inputs.

\subsection{Augmentation}\label{subsec:aug}

We perform two kinds of dataset augmentation on the training set alone, leaving the validation and test sets unchanged. Firstly, we transpose all pieces as many times as possible so that each piece only contains pitches that are within the set of pitches observed in the original dataset, and so that there are no instances of a pitch exceeding the natural range of the voice-type in which it appears. This takes the total number of training examples up to 1,968. We found that transposition makes a significant impact on the model's ability to generalise (Table 1).

We also crudely convert all major pieces to minor, and vice versa, by raising all occurrences of the minor 3rd, 6th and 7th in minor pieces and flattening occurrences of the major 3rd and 6th in major pieces, leaving the 7th raised. This ultimately had negligible impact on model performance and harmed sample quality, while significantly increasing the training time due to effectively doubling the number of samples. We therefore do not consider this an effective technique for dataset augmentation in the context of the JSB chorales. We hypothesise that this weakness may be due to the use of chromaticism and presence of key modulations within samples in the dataset, which may not maintain their stylistic integrity when the overall mode of the piece is converted in the naive manner described.

\section{Models}\label{sec:models}

\subsection{TonicNet}\label{subsec:tonicnet}

TonicNet takes two inputs: the integer corresponding to the previous time-step's class label (X-input) and the integer corresponding to its repetition count (Z-input). These inputs are each converted to one-hot vector representations and by a 256-dimension and 32-dimension embedding respectively. The embedding outputs are then end-concatenated. Both the X embedding and the Z embedding are learned during training.

The concatenated embedding outputs have Variational Dropout [24] applied with a rate of 0.1, before then being input to a three-layer, 256-unit GRU, with dropout applied after each of the first two GRU layers, using a rate of 0.3.

The Z embedding output is reintroduced by end-concatenating with the GRU output, before applying Variational Dropout with a probability of 0.3. The resulting tensor is then fed to a final 98-unit affine layer.

We train TonicNet using a batch size of one for 60 epochs, employing the 1cycle policy [25] with Stochastic Gradient Descent as the optimiser. We begin training with an initial learning rate of 0.008, which is increased to 0.2 over the first 18 epochs. The learning rate is then decreased to 0.0002 over the remaining epochs via cosine annealing. We also cycle momentum inversely to the learning rate between values of 0.8 and 0.95. During training we clip the norm of the gradients to 5, which prevents gradient explosion. Training on the transposed dataset took roughly 3.25hrs on a T4 Tensor Core GPU.

Model hyperparameters, including number of recurrent layers, hidden units and dropout rate, were inspired by [15], and corroborated by initial experimentation. The sizes of both the Z and X embeddings were chosen naively and the effect of varying these has not been determined through experimentation, so there may be room to further tune the parameters of this model and improve results, which we leave to future work.

\begin{figure}
 \centerline{
 \includegraphics[width=1.0\columnwidth]{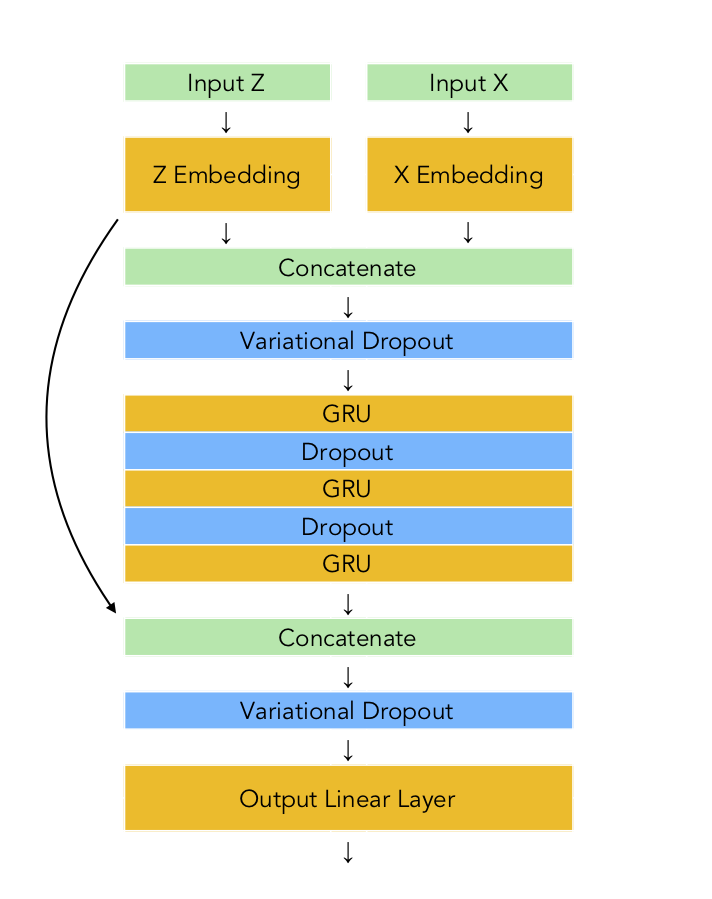}}
 \caption{Diagram of the TonicNet model architecture..}
 \label{fig:tonicnetarch}
\end{figure}

\subsection{Transformer}\label{subsec:transformer}

We use a 5-layer Transformer encoder model. We encode absolute position in the sequence using a fixed sinusoidal position embedding as described in [7], with 256 dimensions. The model also learns a 256-dimension input embedding. The outputs of the two embeddings are end-concatenated. We use 8 attention heads and set the model dimension, D, to 512. The feedforward layer within the encoder module has 1024 units, and dropout is set to 0.1. The hyperparameters of this model are largely derived from the Vanilla Transformer decoder model used as a baseline in [3]. Input is masked to ensure the model can only attend to previous time-steps when making a prediction. Neither instrument labelling or repetition encoding are used when training this model.

We again utilise a batch size of one and employ the 1cycle policy with SGD. In this case we train for 30 epochs, increasing the learning rate from 0.0006 to 0.06 over the first 9 epochs before decreasing to 0.00006 with cosine annealing over the remaining epochs. Momentum is cycled inversely to learning rate between values of 0.8 and 0.95, and the gradient norm is clipped to 5. This model trains faster than TonicNet when measuring time taken per batch, but direct comparison is not possible as it was trained on CPU in our experiments.

\subsection{Common Implementation Details}\label{subsec:com_imp}

In both cases, we derive the maximum learning rate used during training by first performing an LR Range Test [26]. The models are trained using cross-entropy as the objective function, and always see the ground truth values for the previous steps of the sequence when predicting the current step.

While the benchmark dataset contains only notes, and therefore we are arguably more interested in minimising the loss when predicting the pitches of the voices than when predicting chords, we train the model to minimise the average loss across the entire sequence with no bias shown to steps including note predictions. Both models are implemented and trained using the PyTorch library [27].

\section{Evaluation}\label{sec:eval}

\subsection{Quantitative Evaluation}\label{subsec:quant_eval}

The Transformer model is evaluated only on the un-augmented dataset, due to resource constraints. We compare performance when training on just notes of the four voices (SATB), and when training on sequences also including chords (CSATB). TonicNet is tested on a wider range of related tasks including modelling a variety of part combinations, from one part only up to the maximum five parts. We also show the effect of repetition encoding and training set augmentation. Table 1 shows the results of these experiments as measured by validation set loss, and compares performance against two high-performing baselines on the dataset, namely COCONET and Music Transformer.

In Table 1, each model variant shows the parts trained and evaluated on in parentheses, where NLL = Negative Log Likelihood, C = Chords, S = Soprano, A = Alto, T = Tenor and B = Bass. NCL stands for No Chord Loss, indicating that the model was only evaluated on the note predictions, ignoring the loss at time-steps corresponding to chord predictions when averaging NLL. The use of transposition to augment the dataset is denoted by Tr, and MM signifies training set augmentation via major-to-minor key conversion (and vice versa). TonicNet\_Z here indicates the inclusion of repetition encoding Z-inputs when training the model.

The results show a trend whereby training the network to predict more musical information improves overall performance. Both models perform better when trained to predict chords before predicting notes at each 16th-note time-step, as compared with training to predict only the notes. Even when evaluating on the entire CSATB sequence, including the chords in the reported loss, we see that TonicNet\_Z with transposition outperforms Music Transformer, the previous highest-performing ordered model as evaluated on the pure SATB dataset, despite being an order of magnitude smaller. When we ignore loss on chord time-steps, the average NLL is significantly lower, performing better than the upper bound for unordered evaluation of COCONET. The improvement when comparing performance on pitch-wise loss alone is echoed by the Transformer (CSATB) model. From this we can derive that predicting the chords has the impact of significantly improving model confidence when predicting pitches, which is perhaps intuitive.
\newcommand{\astfootnote}[1]{
\let\oldthefootnote=\thefootnote
\setcounter{footnote}{0}
\renewcommand{\thefootnote}{\fnsymbol{footnote}}
\footnote{#1}
\let\thefootnote=\oldthefootnote
}
\begin{table}[]
\begin{tabular}{|l|l|}
\hline
\textbf{Model \& Dataset Variation}                                                                                                                                                                      & \textbf{Validation NLL}                                                                       \\ \hline
\begin{tabular}[c]{@{}l@{}}Transformer (SATB)\\ Transformer (CSATB)\\ Transformer (CSATB, NCL)\end{tabular}                                                                                              & \begin{tabular}[c]{@{}l@{}}0.544\\ 0.503\\ 0.394\end{tabular}                                 \\ \hline
Music Transformer* (SATB)                                                                                                                                                                                 & 0.335                                                                                         \\ \hline
\begin{tabular}[c]{@{}l@{}}COCONET* (SATB, chronological)\\ COCONET* (SATB, orderless)\end{tabular}                                                                                                      & \begin{tabular}[c]{@{}l@{}}0.436\\ $\leq$ 0.238\end{tabular}                                       \\ \hline
\begin{tabular}[c]{@{}l@{}}TonicNet (C) \\ TonicNet (B)\\ TonicNet (S)\\ TonicNet (CS)\\ TonicNet (SB)\\ TonicNet (CSB)\\ TonicNet (SATB)\end{tabular}                                                   & \begin{tabular}[c]{@{}l@{}}0.936\\ 0.716\\ 0.521\\ 0.588\\ 0.555\\ 0.516\\ 0.523\end{tabular} \\ \hline
\begin{tabular}[c]{@{}l@{}}TonicNet\_Z (SATB)\\ TonicNet\_Z (CSATB)\\ TonicNet\_Z (CSATB, Tr)\\ TonicNet\_Z (CSATB, Tr+MM)\\ TonicNet\_Z (CSATB, Tr, NCL)\\ TonicNet\_Z (CSATB, Tr+MM, NCL)\end{tabular} & \begin{tabular}[c]{@{}l@{}}0.497\\ 0.422\\ 0.321\\ 0.317\\ \textbf{0.224}\\ \textbf{0.220}\end{tabular}         \\ \hline
\end{tabular}
\caption{Validation loss on JSB chorales at 16th-note time-steps. }
\label{tab:my-table}
\end{table}

\subsection{Qualitative Evaluation}\label{subsec:qual_eval}

We evaluate the quality of samples from TonicNet via human domain expert analysis. We define a domain expert as someone holding a post-graduate degree in a subject directly related to Western Classical Music, and who has formally studied Bach's chorales as part requirement for obtaining an academic qualification. This is favoured as it allows for a more objective, thorough and strict criticism than a layperson-targeted listening test. We use random sampling to generate chorales from TonicNet, after first selecting a starting minor or major chord at random. Experiments with beam search decoding tended to produce overly-short samples, even when normalising sample probability for length, therefore random sampling according to the output probability distribution is preferred. Stochasticity during beam search has not been subject to experimentation.\let\thefootnote\relax\footnotetext{\textsuperscript{*}Figures reproduced directly from [3]}

We find that TonicNet\_Z (CSATB, Tr) produces the best samples. Voice leading is typically stylistic, especially on a local scale, as is the generated melodic contour, although there is a tendency to diverge from what is clearly the intended phrase within a part for a single 16th-note, usually by a single scale degree or semitone, before then returning, causing an undesirable ornamentation effect.

Harmonisation and harmonic trajectory is also consistently plausible, however there are occasional instances of a phrase which clearly starts in a major key suddenly modulating to a minor key, or vice versa, in a manner that is uncharacteristic to the corpus. The worst generated samples may in fact display poor, overly-chromatic harmonisation and lack stylistic harmonic direction. Some instances of sample weakness may be artefacts of the exposure bias introduced by using teacher forcing when training TonicNet.

The most significant issue detected in samples ultimately relates to phrasing; Bach's chorales feature symmetric phrases, typically two, four or eight bars long, ending in a cadence. Generated pieces have a tendency to feature asymmetry between consecutive phrases, which is not stylistic. Including fermata or other phrase-based information in the modelled sequence could help mitigate this issue, as demonstrated in [5] and [15]. Samples do consistently end on a perfect cadence as expected, and voices never misalign due to a misordering in the generated sequence; rather, each voice clearly completes its phrase, and the duration of each voice's phrase coincides exactly with the others. Samples displaying a range of quality are included in the code repository for fair analysis.

\begin{table}[]
\begin{tabular}{|l|l|l|l|l|}
\hline
\textbf{} & \textbf{Val NLL} & \textbf{Val Acc.} & \textbf{Test NLL} & \textbf{Test Acc.} \\ \hline
Full      & 0.317            & 90.928            & 0.311             & 90.787             \\ \hline
NCL       & 0.220            & 93.468            & 0.214             & 93.419             \\ \hline
\end{tabular}
\caption{Model loss and accuracy when evaluating TonicNet\_Z (CSATB, Tr+MM) on validation and test sets, both when including and ignoring chord predictions (Full versus NCL).}
\label{tab:my-table}
\end{table}

\section{Conclusions}\label{sec:conc}

We first extracted salient features from the existing dataset, in the form of chords and intra-voice token repetition, and then included these extra features among the training inputs. The fact that exposing the model to more features should improve results is unsurprising; more unexpected is that including new features as extra elements within the series being modelled should dramatically enhance performance. Furthermore, it was noted that confidence when predicting pitches is much higher if the model is first tasked with predicting chords. This suggests a worthwhile area for further research is to improve confidence when predicting the chords, and we conjecture that a method to achieve this may be to include yet more related features in the sequences themselves, such as fermata information or a representation of floating tonality, given our findings.

State-of-the-art validation loss on the JSB chorales dataset was achieved with a variation of TonicNet and effective dataset augmentation, and we demonstrated that despite this there are still some specific weaknesses in sample quality which other approaches have mitigated. We also noted the superior human interactability of COCONET and DeepBach, which we believe gives those proposals a greater potential for real-world application. However, the findings presented in this paper could be applicable to a wide range of approaches to statistical modelling of polyphonic music, and their merit was demonstrated on two such approaches.

We also surveyed the effects of serialising music by splitting notes across a fine-resolution temporal grid. While the benefit of this in terms of data augmentation was noted, we also presented weaknesses relating to true rhythmic integrity in the case of repeated note boundaries, and vast extension of sequence length which has the effect of increasing both training and sampling time. Ultimately we would like to move towards utilising encodings including true rhythmic duration, with the aim of being able to train more general polyphonic music models that are not confined to a temporal grid, and therefore to styles of music whose rhythmic resolution can be encompassed by this grid. It is not viable to simply serialise music into increasingly finer-resolution rhythmic units in order to accommodate datasets which include some occurrences of tuplet durations, for example, as this continues to extend the overall sequence length. Future work will explore solutions to this drawback.

\section{Acknowledgements}\label{sec:ack}

We thank Jack Kemp for his helpful suggestions and feedback, all of which served to improve the earliest drafts of this work.


\end{document}